\documentclass[aps,prl,twocolumn,showpacs,preprintnumbers,superscriptaddress]{revtex4-1}
\usepackage{amsfonts}
\usepackage{graphicx}
\usepackage{dcolumn}
\usepackage{bm}

\newcommand{\nc}{\newcommand}
\nc{\be}{\begin{equation}}
\nc{\ee}{\end{equation}}
\nc{\bea}{\begin{eqnarray}}
\nc{\eea}{\end{eqnarray}}
\nc{\bean}{\begin{eqnarray*}}
\nc{\eean}{\end{eqnarray*}}
\nc{\mb}{\mbox}
\nc{\rnc}{\renewcommand}
\nc{\vk}{\mb{\bf k}}
\nc{\vp}{\mb{\bf p}}
\nc{\vn}{\mb{\bf n}}
\nc{\vq}{\mb{\bf q}}
\nc{\rr}{\mb{\bf r}}
\nc{\vz}{\hat {\mb{\bf z}}}
\nc{\vj}{\mb{\boldmath$j$}}
\nc{\vg}{\mb{\boldmath$g$}}
\nc{\x}{\mb{\boldmath$x$}}
\nc{\A}{\mb{\boldmath$A$}}
\nc{\va}{\mb{\boldmath$a$}}
\nc{\vs}{\mb{\boldmath$\sigma$}}
\nc{\vpi}{\mb{\boldmath$\pi$}}
\nc{\nab}{\nabla}
\nc{\X}{\sf x}

\begin{document}

\title{Quantum Hall to charge-density-wave phase transitions in ABC-trilayer
graphene}
\author{Yafis Barlas}

\affiliation{Department of Physics and Astronomy, University of California,
Riverside, CA 92521}

\author{R. C\^{o}t\'{e}}
\affiliation{D\'{e}partement de physique, Universit\'{e} de Sherbrooke, Sherbrooke
(Qu\'{e}bec), Canada, J1K 2R1}

\author{Maxime Rondeau}
\affiliation{D\'{e}partement de physique, Universit\'{e} de Sherbrooke, Sherbrooke
(Qu\'{e}bec), Canada, J1K 2R1}


\begin{abstract}
ABC-stacked trilayer graphene's chiral band structure results in three ($n=0,1,2$) 
Landau level orbitals with zero kinetic energy. This unique feature has 
important consequences on the interaction driven states of the 12-fold degenerate 
(including spin and valley) $N=0$ Landau level. In particular, at many filling factors 
$\nu_{T} =\pm5,\pm4,\pm2,\pm1$ a quantum phase transition from a quantum Hall liquid 
state to a triangular charge density wave occurs as a function of the single-particle 
induced LL orbital splitting $\Delta_{LL}$. Experimental signatures of this phase transition
are also discussed.
\end{abstract}

\date{\today}
\maketitle

\begin{figure}[tbp]
\includegraphics[width=3.70in,height=2.0in]{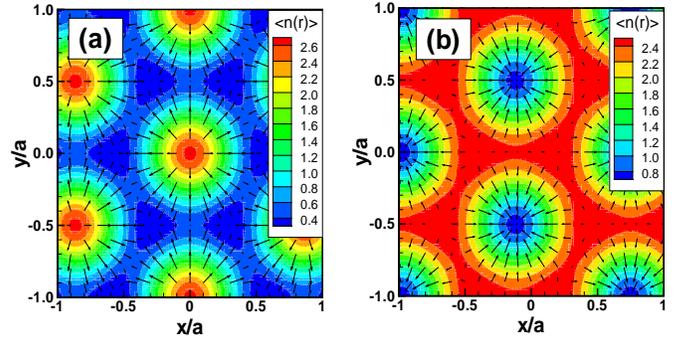}
\caption{(Color online) Electronic density and orientation of the in-plane
electric dipole field for the triangular crystal at filling factor (a) $%
\protect\nu =-5,-2,1,4$ and (b) $\protect\nu =-4,-1,2,5.$The electrons for
(a) and holes for (b) crystallize in a triangular pattern with a lattice
constant $a=\protect\sqrt{4\protect\pi /\protect\sqrt{3}}l_{B}$. The
crystal state is characterized by a superposition of three LL orbitals $%
n=0,1,2$ pseudospin which manifests itself as a periodic density modulation $%
n\left( \mathbf{r}\right) $ (see text for details).}
\label{crystal}
\end{figure}

\emph{Introduction}-- When a group of partially filled Landau levels (LLs)
with internal flavor indices are nearly degenerate, electron-electron
interactions can lead to spontaneous broken symmetries which induce
quasiparticle gaps and hence interaction driven integer quantum Hall (IQH)
effects~\cite{GirvinAHM}. In semiconducting two-dimensional electron systems
(2DES) where Zeeman splitting is negligible and the LLs consist of narrowly
spaced doublets, interactions lead to spontaneous electron spin polarization~%
\cite{sondhi} accompanied by a large exchange enhanced spin gap. 
More recently interaction driven broken symmetry states at integer fillings
have been observed in both graphene~\cite{qhf_graph,kenallan} and bilayer
graphene~\cite{ybqhf_bg,qhf_bg} which in most cases exhibits a set of
four-fold degenerate LLs. At many filling factors this four-fold degeneracy,
which is due to the spin and valley degree of freedom, is spontaneously
broken resulting in charge gaps and hence incompressible quantum Hall (QH)
states. Bilayer graphene's chiral bands lead to an additional degeneracy
doubling of the zero-energy LL associated with the $n=0$ and $n=1$ LL
orbitals~\cite{falkomcann}, leading to an eight-fold degenerate LL. This 
additional LL degeneracy also results in interaction driven QH effects due 
to an exchange induced gap between the LL pseudospins~\cite{ybqhf_bg,anamextcond}. \newline
\begin{figure}[t]
\begin{center}
\includegraphics[clip,width=3.00in,height=2.15in]{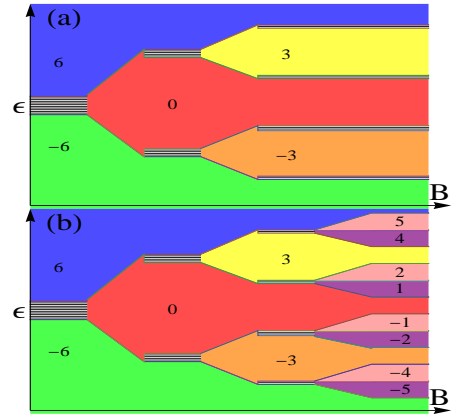}
\end{center}
\caption{(Color online) Schematic representation of the interaction induced energy 
gaps as a function the magnetic field for interaction driven states in ABC-trilayer 
graphene's 12-fold degenerate LL, the numbers represent the Hall conductivity in units of $e^2/h$. 
For (a) $\Delta_{LL} < \Delta^{(c)}_{LL}$ the presence of insulating triangular CDW states for 
$\nu = \pm 5, \pm4, \pm 2 ,\pm 1 $ results in Hall plateaus only at $\sigma_{xy} = \pm 6,
\pm 3, 0 (e^2/h)$. For (b) $\Delta_{LL} > \Delta^{(c)}_{LL}$ interaction driven QH states 
appear at all intermediate integer fillings with the sequence 
determined by Hund's rule behavior (see text for details).}
\label{figone}
\end{figure}
\indent At relevant energy scales few layer graphene with different
stacking sequences, referred to as chiral 2DES (C2DES), exhibit electronic
properties distinct from semiconducting 2DES~\cite{HongkiAHM}. The
difference in the LL structure of C2DES is most evident in the zero-energy
LL: for certain stacking sequences (such as AB, ABC, ABCB, etc.) multiple LL
orbitals are degenerate~\cite{QHgrapreview}. The role of interactions on the
ground state when multiple LL orbital degrees of freedom are degenerate
remains relatively unexplored. In this Letter we show that, unlike bilayer graphene, 
at certain filling factors interaction induced QH states are unstable in 
ABC-stacked trilayer graphene's 12-fold degenerate zero-energy LL. This is due to 
the presence of a degenerate set of triplet ($n=0,1,2$) LL orbitals.
While the QH liquid state is stable for $\nu =-3,3,6$ the nature of the
ground states at $\nu =\pm5,\pm4,\pm2,\pm1$ depends on the ratio of a single particle induced LL
orbital splitting $\Delta _{LL}$ and electron-electron interactions $%
e^{2}/\varepsilon l_{B}$ ($l_{B}=\sqrt{hc/eB}$ is the magnetic length). In
particular, a quantum phase transition occurs from a triangular charge
density wave (CDW) to a translationally invariant QH liquid state as
$\Delta_{LL}$ is increased beyond a critical value $\Delta _{LL}^{\left( c\right) }
$\cite{footnoteexp}. This transition is first order in the liquid-solid universality class.
The crystal state exhibits coherence between LL orbitals which leads to 
vortex textures of in-plane electric dipoles at each lattice site.
For moderately disordered samples this phase transition should
be characterized by a Hall plateau transition as a function of $\Delta_{LL}$
at a \textit{fixed} filling factor $\nu$. Another consequence is that the Hall conductivity 
for $\Delta_{LL} < \Delta^{(c)}_{LL}$ will correspond to the \textit{adjacent} 
interaction driven IQH plateau, for example, at $\nu = -5,-4$, $\sigma_{xy} = -6e^2/h$ 
(as shown in Fig.~\ref{figone}a).\newline
\indent\emph{ABC-trilayer graphene}-- ABC-stacked graphene trilayers are
chiral generalization of monolayer and Bernal-stacked bilayer graphene.
ABC-stacked trilayer is a particular stacking of three graphene layers each
with inequivalent sublattices $A_{i}$ and $B_{i}$ arranged in following
sequence: one of the two-carbon atom sites in both the top and bottom layer $%
B_{1}(A_{3})$ has a different near-neighbor carbon atom site in the middle
layer $A_{2}(B_{2})$, which leaves one-carbon atom site in the top and
bottom layers $A_{1}(B_{3})$ without a near-neighbor in the middle layer.
Interlayer hopping on adjacent layer near-neighbor carbon atom sites leads to
the formation of high-energy dimer bands which push the electron energy away
form the Fermi surface leaving one low-energy sublattice site per $\pi $%
-carbon orbital in the outermost layers. Because the hopping between the
low-energy sites via intermediate high-energy states is a three-step
process, it turns out that in the presence of an external magnetic field
three LL orbitals $(n=0,1,2)$ have zero kinetic energy. The wavefunctions
of the opposite ${\mathbf{K}}({\mathbf{K}}^{\prime })$ valleys in the zero-energy LL
are localized in the top(bottom) layers, hence layer and valley labels are
equivalent. This simplified two-band model for the ABC-stacked graphene
trilayer which includes only nearest-neighbor interlayer and intralayer
hopping terms is valid at large external magnetic fields~\cite{inprep}.
The effects of an external potential difference between the top and bottom
layers $\Delta _{V}$ along with remote next-nearest interlayer hopping terms
on the zero-energy LL, not included in our simplified model, can be
incorporated by introducing a symmetric LL gap $\Delta _{LL}$~\cite%
{footnoteone} between the degenerate zero-energy triplet set $(n=0,1,2)$ of
LL orbitals. The effect of these terms on $\Delta _{LL}$ will be reported
elsewhere~\cite{inprep}. \newline
\indent\emph{Mean field theory}-- Trilayer graphene's zero-energy LL is a
direct product of two $S=1/2$ doublets corresponding to spin and valley
(layer) pseudospin along with $S=1$ LL orbital pseudospin which is
responsible for the new physics discussed in this Letter. Working in the
Landau gauge $\mathbf{A}=(0,Bx,0),$ the Hartree-Fock (HF) mean-field
Hamiltonian, projected onto the zero-energy LL of ABC-stacked trilayer,
consists of single particle pseudospin splitting terms along with direct and
exchange interaction contributions: $\mathcal{H}_{HF} = \mathcal{H}_{sp} + 
\mathcal{H}_{int}$, with  
\begin{eqnarray}
\label{interhamp}
\frac{\mathcal{H}_{int}}{N_{\phi }} &=&\sum_{\mathbf{q} \neq 0,r,s}H_{n_{1}n_{2}n_{3}n_{4}}^{\alpha \beta }(%
\mathbf{q})  \langle \Delta
_{\alpha s;\alpha s}^{n_{1}n_{4}}(\mathbf{q})\rangle \Delta _{\beta r;\beta r}^{n_{2} n_{3}}(-\mathbf{q})\nonumber  \\
&-&\sum_{\mathbf{q},r,s}X_{n_{1}n_{2}n_{4}n_{3}}^{\alpha \beta }(\mathbf{q})\langle \Delta
_{\alpha s;\beta r}^{n_{1} n_{4}}(\mathbf{q})\rangle \Delta _{\beta r;\alpha s}^{n_{2} n_{3}}(-\mathbf{q}),
\end{eqnarray}%
(repeated indices are summed over) where $n_{i}=0,1,2$ are the LL orbital
indices, $\alpha ,\beta =t(b)$ label the top (bottom) layers (${\mathbf K}({\mathbf K}')$ valleys)
and $r,s=1(-1)$ label the $\uparrow (\downarrow )$ spins, $N_{\phi }=A/(2\pi
l_{B}^{2})$ is the degeneracy of a single LL and momentum space guiding
center density is , 
\begin{equation}
\Delta _{\alpha s;\beta r}^{nn^{\prime }}(\mathbf{q})=\frac{1}{N_{\phi }}
\sum_{X,X^{\prime }}c_{n X \alpha s}^{\dagger }c_{n^{\prime } X^{\prime
} \beta r}e^{-i\frac{q_{x}}{2}(X^{\prime }+X)}\delta _{q_{y}l_{B}^{2},X-X^{\prime }},
\label{coherences}
\end{equation}%
where $c_{nX\alpha s}^{\dagger }(c_{nX\alpha s})$ are the respective $n^{th}$
LL creation and annihilation operators at a given guiding center $X$ in the
Landau gauge with layer and spin index $\alpha ,s$. The effects of the single-particle
pseudospin splitting field are encoded in 
\begin{equation}
\frac{\mathcal{H}_{sp}}{N_{\phi}}=\bigg(n\Delta _{LL}-\xi _{\alpha }\frac{\Delta _{V}}{2}-s\frac{%
\Delta _{Z}}{2} \bigg)\rho _{\alpha s}^{n},
\end{equation}
where $\rho _{\alpha s}^{n}=\Delta _{\alpha s;\alpha s}^{n,n}(\mathbf{q}%
=0)=\sum_{X}\langle c_{nX\alpha s}^{\dagger }c_{nX\alpha s}\rangle $ is the
total guiding center density for the $n^{th}$ LL with valley and spin index $%
\alpha ,s$ and  $\Delta _{LL}$ is the LL splitting induced by the remote
interlayer hopping terms and the external potential difference $\Delta _{V}$
with $\xi _{t(b)}=1(-1)$, $\Delta _{Z}=g\mu _{B}B$ is the Zeeman splitting
and all energies are measured in units of $e^{2}/\varepsilon l_{B}$. In (\ref%
{interhamp}) the Hartree-field captures the valley (layer) dependent electrostatic
contributions: 
\begin{equation}
H_{n_{1} n_{2} n_{3} n_{4}}^{\alpha \beta }(\mathbf{q},d)=\frac{1}{2\pi
l_{B}^{2}}v_{\mathbf{q}}^{\alpha \beta }F_{n_{1} n_{4}}(\mathbf{q}%
)F_{n_{2} n_{3}}(-\mathbf{q}),
\end{equation}%
and the exchange contribution is captured by: 
\begin{eqnarray}
X_{n_{1} n_{2} n_{3} n_{4}}^{\alpha \beta }(\mathbf{q},d)&=&\frac{1}{L^{2}}%
\sum_{\mathbf{p}}v_{\mathbf{p}}^{\alpha \beta }F_{n_{1} n_{4}}(\mathbf{p}%
)F_{n_{2} n_{3}}(-\mathbf{p})\nonumber \\ & &  
\times e^{i(\mathbf{p}\times \mathbf{q})l_{B}^{2}},
\end{eqnarray}%
where $v_{\mathbf{q}}^{tt}=v_{\mathbf{q}}^{bb}=2\pi e^{2}/\varepsilon |%
\mathbf{q}|$ and $v_{\mathbf{q}}^{tb}=v_{\mathbf{q}}^{bb}e^{-qd}$~\cite{footnote} is the
Fourier transform of electron-electron interactions in the same and
different layers (valleys) and $d=0.667\mathrm{nm}$ is the interlayer
separation. The form factors $F_{nn^{\prime }}(\mathbf{q})$ capture the
character of the three different LL orbitals and are given by: 
\begin{equation}
F_{n n^{\prime }}(\mathbf{q})=\sqrt{\frac{n^{\prime }!}{n!}}\bigg(\frac{%
i|q|e^{i\theta _{q}}l_{B}}{\sqrt{2}}\bigg)^{n-n^{\prime }}L_{n^{\prime
}}^{n-n^{\prime }}\bigg(\frac{(ql_{B})^{2}}{2}\bigg)e^{\frac{-(ql_{B})^{2}}{4%
}},
\end{equation}%
for $n^{\prime }\leq n$, where $L_{n}^{\alpha }(x)$ is the generalized
Laguerre polynomial, $\theta _{q}=\tan ^{-1}(q_{y}/q_{x})$ and the form
factors satisfy $F_{nn^{\prime }}(\mathbf{q})=[F_{n^{\prime }n}(-\mathbf{q}%
)]^{\ast }$.\newline
\indent First we look for translationally invariant solutions of the density
matrix $\langle \Delta _{\alpha s;\alpha ^{\prime } s^{\prime }}^{n,n^{\prime }}(%
\mathbf{q}=0)\rangle$. The density matrix must be determined by self-consistently
occupying the lowest energy eigenvectors of $\mathcal{H}_{HF}$. For a
uniform density order parameter the only contribution to the Hartree term is 
$E_{H}=(e^{2}/\varepsilon l_{B})(d/2l_{B})$, which captures the charging
capacitance energy of the outermost layers of ABC-trilayers. The filling
order of the 12-fold degenerate zero-energy LL proceeds in integer
increments starting from the filling factor $\nu =-6$ and follows a Hund's
rule behavior: first maximize spin polarization, then maximize the layer
pseudospin polarization to the greatest possible extent and finally maximize
the LL polarization to the greatest extent allowed by the first two rules (see Fig.~\ref{figone}b). \newline
\begin{figure}[t]
\begin{center}
\includegraphics[clip,width=3.375in,height=2.25in]{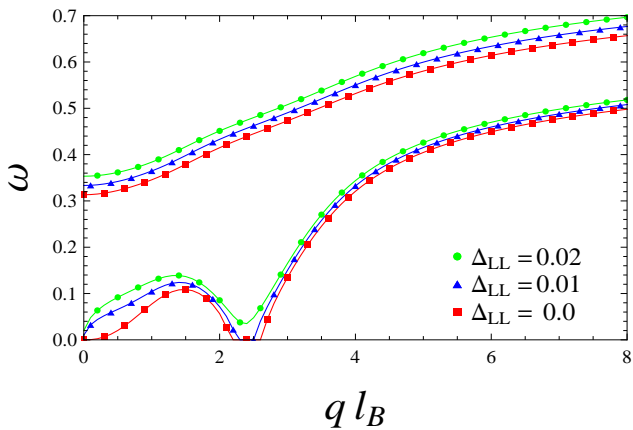}
\end{center}
\caption{(Color online) Collective mode dispersion $\protect\omega _{q}$ for $%
\protect \nu =-5,-2,1,4$ in units of interaction strength $e^{2}/%
\protect\varepsilon l_{B}=11.2(B[T])^{1/2}$ meV as a function of $ql_{B}$ at
different values of the LL splitting $\Delta _{LL}$ at a magnetic field of $%
B=10$ Tesla. Charge density wave instability is accompanied by a softening
of the magneto-roton minima for $\Delta _{LL}\leq \Delta _{LL}^{(\ast) }$.
The gapped mode corresponds to the hard-direction in LL pseudospin space and
is not essential for our discussion (see text for details).}
\label{figtwo}
\end{figure}
\indent
For balanced ABC-trilayers ($\Delta _{V}=0$) at $\nu = -3,3$ the valley(layer)
pseudospin exhibits spontaneous inter-valley(layer) coherent state similar to that observed 
in conventional 2DEGs~\cite{spielman}. The $\nu =0 $ state is spin polarized as the 
exchange gap associated with spin polarization is larger than the exchange gap associated 
wiht the inter-layer coherent state.
However, since the energy scales associated with Zeeman splitting and the 
ratio $d/l_{B}$  are small compared to typical interaction
energies, the valley and spin exchange gaps compete closely which could
alter the Hund's rule behavior (for example leading to density wave states
rather than spin-polarized states at $\nu =0$). Another consequence of a small 
value of $d/l_{B}$ is that for a small value of $\Delta _{V} (\geq (\pi /8)^{1/2}(d/l_{B})^{2} 
e^{2}/\varepsilon l_{B}$) the inter-layer coherent state switches to a valley 
polarized state. \newline
\indent
Hund's rule dictate that the topmost LLs at $\nu =-5,-2,1,4$ have $n=0$ LL
orbital occupation whereas at $\nu=-4,-1,2,5$ the $n=1$ LL orbital is occupied.
This preference of LL orbital occupation is associated with a better exchange energy 
of the more localized $n=0$ LL orbital wavefunction along with the absence of nodes~\cite{ybqhf_bg}. 
The new physics discussed in this Letter is primarily associated with the 
LL pseudospin ordering at $\nu \neq -3,0,3,6$. In the next section we show that for $%
\Delta _{LL}<\Delta _{LL}^{(\ast) }$ the translationally invariant uniform
liquid QH states at $\nu =-5,-2,1,4$ are unstable to the LL orbital
fluctuations. This instability is due to strong charge fluctuations which lead
to a softening of the collective mode dispersion as a function of 
$\Delta_{LL}$.\newline
\indent
\emph{Charge density wave instability}---We now focus on the
collective excitations of a filled QH liquid at filling factors $\nu
=-5,-2,1,4$, freezing spin, and valley (layer) but keeping the ($n=0,1,2$)
LL orbital degrees of freedom. To further simplify the discussion and 
concentrate on the LL pseudospin fluctuations, we also assume SU(4)-invariant 
spin and valley (layer) independent interactions.~\cite{footnotetwo} In this case the SU(4)-flavor index is
spontaneously broken and quantized LL pseudospin fluctuation consists of
pseudospinor rotations which mix $n=0$ with $n=1$ and $n=2$ LL orbitals leading 
to two branches of collective mode excitations. The
collective mode excitations which we evaluate below are primarily influenced
by the orbital dependence of the microscopic Hamiltonian (\ref{interhamp}).
A detailed account of the collective excitations for $d\neq 0$ will be
presented elsewhere~\cite{inprep}. \newline
\indent
One physically transparent way of performing these collective mode
calculations is to construct a fluctuation action in which each transition
has canonically conjugate density $\rho $ and phase $\varphi $ components
corresponding to the real and imaginary parts of the final state component
in the fluctuating spinor. 
The fluctuation action 
\begin{equation}
\mathcal{S}[\mathbf{\rho },\mathbf{\varphi }]=\mathcal{S}_{B}[\mathbf{\rho },%
\mathbf{\varphi }]-\int d\omega \int d^{2}q\;\mathcal{E}[\mathbf{\rho },%
\mathbf{\varphi }],  \label{action}
\end{equation}%
contains a Berry phase term $\mathcal{S}_{B}$~\cite{auerbach} 
\begin{equation}
\mathcal{S}_{B}[\mathbf{\rho },\mathbf{\varphi }]=\int d\omega \int d^{2}q%
\left[ \frac{1}{2}\mathbf{\rho }_{-\mathbf{q}}^{\dag }\mathcal{D}\mathbf{%
\varphi }_{\mathbf{q}}-\mathbf{\varphi }_{-\mathbf{q}}^{\dag }\mathcal{D}%
^{\dagger }\mathbf{\rho }_{\mathbf{q}}\right] ,
\end{equation}%
where $\mathcal{D}=-i\omega \mathbb{I}_{2\times 2}$ and $\mathbf{\rho }_{%
\mathbf{q}}=(\rho _{1,\mathbf{q}},\rho _{2,\mathbf{q}})$ with $\rho _{n,%
\mathbf{q}}$ corresponding to the charge fluctuation of the $n=0$ to the $n^{th}$
orbital, with a similar definition for the phase variable $\varphi _{\mathbf{%
q}}$. The energy functional, $\mathcal{E}[\mathbf{\rho },\mathbf{\varphi }]$
is closely related to the discussion in the preceding section 
\begin{equation}
\mathcal{E}[\mathbf{\rho },\mathbf{\varphi }]=\frac{1}{2}\left[ \mathbf{\rho 
}_{-\mathbf{q}}^{\dag }(\Gamma _{q}+\Lambda _{q})\mathbf{\rho }_{\mathbf{q}}+%
\mathbf{\varphi }_{-\mathbf{q}}^{\dag }(\Gamma _{q}-\Lambda _{q})\mathbf{%
\varphi }_{\mathbf{q}}\right] .
\end{equation}%
Here, $\Gamma _{q}$ and $\Lambda _{q}$ are $2\times 2$ Hermitian matrices
which capture the energy cost of small pseudospinor fluctuations and can be
evaluated explicitly. 
The matrix elements of $\Gamma _{q}$ can $\Lambda _{q}$ can be expressed in
terms of direct and exchange matrix elements: 
\begin{eqnarray}
\gamma _{ij} &=&a_{i}\delta _{ij}+(-1)^{i+j}H_{i0j0}(|\mathbf{q}%
|,0)-X_{i00j}(\mathbf{q},0),  \nonumber \\
\lambda _{ij} &=&(-1)^{i+j}H_{i0j0}(|\mathbf{q}|,0)-X_{00ij}(\mathbf{q},0),
\end{eqnarray}%
where $a_{1}=\Delta _{LL}+1/2\sqrt{\pi /2}$ and $a_{2}=2\Delta _{LL}+5/8%
\sqrt{\pi /2}$ are wavevectors independent contributions measured in units of 
$e^2/\varepsilon l_{B}$. \newline
\indent The collective mode dispersion plotted in Fig.~\ref{figtwo} indicates 
that the uniform QH liquid state is unstable for $\Delta_{LL} \leq \Delta^{(\ast)} = 0.015 
e^2/(\varepsilon l_{B})$. The gapped mode in Fig.~\ref{figtwo} which corresponds to some wavevector
dependent linear combination of the orbital pseudospin fluctuations ($\rho
_{n,\mathbf{q}},\varphi _{n,\mathbf{q}}$) reflects the difference in the
exchange energy cost of the LL orbitals. The \textquotedblleft
soft-mode\textquotedblright\ indicates the instability of the uniform QH liquid 
at $q_{0}l_{B}\sim 2.37$ which is due to the charge
fluctuations of the QH liquid as we show next. \newline 
\indent In order to analyze this instability and its associated consequences
it is convenient to expand the energy functional of the \textquotedblleft
soft-mode\textquotedblright\ density $\rho _{-,\mathbf{q}}$ near $q\sim q_{0},
$ 
\begin{eqnarray}
\mathcal{E}[\rho _{-,\mathbf{q}}] &=&\left[ \alpha (\Delta _{LL})+\xi
(\Delta _{LL})(q-q_{0})^{2}\right] \rho _{-,\mathbf{q}}^{2}
\label{enefuncsoftmode} \\
&+&\beta (\Delta _{LL})\rho _{-,\mathbf{q}_{1}}\rho _{-,\mathbf{q}_{2}}\rho
_{-,\mathbf{q}_{3}}\delta (\mathbf{q}_{1}+\mathbf{q}_{2},\mathbf{q}%
_{3})+\cdots   \nonumber
\end{eqnarray}%
where $\alpha (\Delta _{LL})=1.24(\Delta _{LL}-\Delta _{LL}^{\left( \ast
\right) })$ with $\xi (\Delta _{LL})=0.29-0.14\Delta _{LL}$
and $\beta (\Delta _{LL})$ is the coefficient of the third order term which is not prohibited by any
symmetries. For $\beta =0$ (\ref{enefuncsoftmode}) exhibits a
Brazovskii-like instability leading to a fluctuation-induced
isotropic-smectic transition to a unidirectional LL-pseudospin density wave
pattern~\cite{Brazovskii}. However, the presence of a finite $\beta $
preempts this transition to a triangular two-dimensional CDW 
(or crystal)\cite{alexandermctague} with LL orbital pseudospin
textures.\newline 
\indent 
This instability should be present whenever ($n \geq 2$) LL orbitals are degenerate. The effective interaction 
within the projected $n^{th}$ LL orbital $v_{n}(q) = v_{q} [L_{n} (q^2 l_{B}^2/2)]^2 e^{-q^2 l_{B}^2/2}$
has minimas corresponding to zeros of the Laguerre polynomials. Due to this the Hartree 
contribution vanishes at these finite-$q$ wavevectors and since the exchange energy 
is positive for all wavevectors, $E_{HF}$ becomes negative~\cite{fogler}. The near degeneracy of 
$n=0,1,2$ LL orbitals allows electrons access to lower energy by forming coherent superposition 
of LL orbitals for a range of finite-$q$ wavevectors, leading to a CDW instability.
As $\Delta_{LL}$ is increased from below this gain in energy competes with
the single particle contribution and the uniform QH state becomes energetically more 
favorable resulting in a transition. \newline
\indent\emph{Triangular Charge Density Wave ($\Delta _{LL}<\Delta _{LL}^{c}$)%
}---For $\Delta _{LL}<\Delta _{LL}^{\left( c\right) }$ with $\Delta
_{LL}^{\left( c\right) }=0.095e^{2}/\varepsilon l_{B}>\Delta _{LL}^{\left(
\ast \right) }$ at $\nu =-5,-2,1,4,$ the HF solution with the lowest energy
is a triangular crystal of electrons with one electron per lattice site as
shown in Fig. \ref{crystal}~\cite{footnotethree}. In this crystal state, all three orbitals $%
n=0,1,2$ are occupied in contrast with the QH liquid state where only $n=0$
is fully occupied. As $\Delta _{LL}\rightarrow \Delta _{LL}^{\left( c\right)
}$ from below, the total population $\left\langle \Delta _{1,1}\right\rangle
+\left\langle \Delta _{2,2}\right\rangle $ of the higher-energy orbitals
decreases but stays finite until $\Delta _{LL}^{\left( c\right) }$ where it
drops discontinuously to zero indicating a first order transition to the
liquid state with $\left\langle \Delta _{0,0}\right\rangle =1$. At each
crystal site, the electron is in a superposition of the three orbital states 
$n=0,1,2.$ Since the wavefunction profile of the three orbitals is
different, this superposition leads to the triangular CDW pattern
in the total electron density $n\left( \mathbf{r}\right) $ shown in Fig. \ref%
{crystal} (a). When two orbital are
filled (at $\nu =-4,-1,2,5$ for example), the ground state at small $\Delta_{LL}$ is a
crystal of holes instead of electrons (see Fig\ref%
{crystal} (b) ). \newline
\indent The CDW state is characterized by a periodic modulation of $%
\left\langle \Delta _{n,m}(\mathbf{q})\right\rangle (n\neq m)$ defined in
Eq. (\ref{coherences}). This leads to the presence of an in-plane electric
dipole field~(a similar situation also occurs in bilayer graphene\cite%
{shizuya}):%
\begin{eqnarray}
d_{i=x,y}\left( \mathbf{q}\right)  &=&-\sqrt{2}el_{B}e^{-q^{2}l_{B}^{2}/4}%
\left\langle \rho _{i}^{\left( 0,1\right) }\left( \mathbf{q}\right)
\right\rangle  \\
&&-2el_{B}e^{-q^{2}l_{B}^{2}/4}\left( 1-q^{2}l_{B}^{2}/4\right) \left\langle
\rho _{i}^{\left( 1,2\right) }\left( \mathbf{q}\right) \right\rangle , 
\nonumber
\end{eqnarray}%
where $\rho _{x}^{\left( n,m\right) }=\left( \Delta _{n,m}+\Delta
_{m,n}\right) /2,\rho _{y}^{\left( n,m\right) }=\left( \Delta _{n,m}-\Delta
_{m,n}\right) /2i$ (with $n>m$), the energy of the CDW state in the presence of an external
electric field is $-\int d\mathbf{r}\left( \mathbf{d}\left( \mathbf{r%
}\right) \cdot \mathbf{E}\left( \mathbf{r}\right) \right).$  Fig. \ref%
{crystal}(a)-(b) also show the orientation of the in-plane electric dipole field
around each crystal site which exhibits vortex textures in which the
dipole vectors rotate by $2\pi .$ The topological properties of this crystal
state will be characterized elsewhere~\cite{inprep}. \newline
\indent The degeneracy of $n=2$ LL orbital is crucial for the existence of
the crystal state. This situation is somewhat reminiscent of the various crystal
phases that appear in conventional 2DES at fractional fillings of higher LLs 
$n\geq 2$~\cite{fogler,bubblestripes}. The surprising result here is that 
similar electronic crystal phases appear at total {\em integer} filling factors 
in ABC-stacked trilayer graphene and are sensitive to the competition of interactions 
and single particle terms induced by remote hopping and layer imbalance. 
This triangular CDW state would likely be pinned by disorder leading to
insulating behavior of the uppermost LL thus exhibiting quantized Hall
conductivity at a value corresponding to the \textit{adjacent} interaction
driven IQH plateau (see Fig.~\ref{figone}a). The triangular CDW state should also be visible through its
pinning mode in microwave absorption experiments.

Acknowledgements: The authors would like to thank C. M. Varma and A. H. MacDonald for 
many fruitful discussions and comments on the manuscript. R. C\^{o}t\'{e} was supported 
by  Natural Sciences and Engineering Research Council of Canada (NSERC), Compute Canada and 
Calcul Qu\'{e}bec.


\begin{thebibliography}{99}
\bibitem{footnoteexp}
$\Delta_{LL}$ can be increased by applying an external potential difference 
$\Delta_{V}$ between the outermost layers.
\bibitem{footnoteone} In this Letter we require a symmetric $0 \leq \Delta_{LL}
< 0.3 e^2/\varepsilon l_{B}$. The phase diagram for $\Delta_{LL} <0$ and for
an asymmetric LL orbital splitting will be reported elsewhere~\cite{inprep}.
\bibitem{footnotetwo} This approximation is well justified since the layer
separation is merely a fraction of a nanometer.
\bibitem{footnotethree}
Laughlin-like correlated states cannot be captured in our HF approximation and require an exact
diagonalization approach which remains technically challenging for the case
of multiple LL orbitals.
\end{thebibliography}
\end{document}